# HistoSeg : Quick attention with multi-loss function for multi-structure segmentation in digital histology images

Saad Wazir
*National University of Sciences and Technology (NUST), Islamabad, Pakistan.*
*swazir.mscs18seecs@seecs.edu.pk*

Muhammad Moazam Fraz
*National University of Sciences and Technology (NUST), Islamabad, Pakistan.*
*moazam.fraz@seecs.edu.pk*

*Abstract*—Medical image segmentation assists in computer-aided diagnosis, surgeries, and treatment. Digitize tissue slide images are used to analyze and segment glands, nuclei, and other biomarkers which are further used in computer-aided medical applications. To this end, many researchers developed different neural networks to perform segmentation on histological images, mostly these networks are based on encoder-decoder architecture and also utilize complex attention modules or transformers. However, these networks are less accurate to capture relevant local and global features with accurate boundary detection at multiple scales, therefore, we proposed an Encoder-Decoder Network, Quick Attention Module and a Multi Loss Function (combination of Binary Cross Entropy (BCE) Loss, Focal Loss Dice Loss). We evaluate the generalization capability of our proposed network on two publicly available datasets for medical image segmentation MoNuSeg and GlaS and outperform the state-of-the-art networks with 1.99% improvement on the MoNuSeg dataset and 7.15% improvement on the GlaS dataset. Implementation Code is available at this link: https://bit.ly/HistoSeg

## I. INTRODUCTION

In computer-aided diagnosis, a key role is played by segmentation of histological images in the number of applications, Detection, Segmentation, and Staging of objects in histological images [17, 7, 5, 18, 4, 19] like Glands, Benin, and Malignant nuclei or other biomarkers helps in monitoring, diagnosis, and treatment of patients. It also helps in surgery which is dependent on image-guided principles. Therefore, in the modern era, a variety of methods like creating state-of-the-art and intelligent methods for the segmentation of histological images are very popular. In the past few years, Deep Convolution Neural Networks (DCNN) took place of the most accurate and intelligent methods for the segmentation of histological images, and also the performance and accuracy of histological image segmentation is greatly increased.DCNN models which are specially designed for histological image segmentation like U-Net[16], MSRF-Net[22], U-Net++[27], and MedT[9] achieves greater performance and accuracy on very difficult datasets and prove that DCNN are very successful to learn the features to segment organs, nuclei, and other objects in histological images.

In segmentation, Encoder-Decoder networks show massive performance boosts like U-Net and DeepLabv3+ [3]. Their Encoder-Decoder networks contain convolutions, pooling, and atrous convolutions with skip connections which make them simple and powerful as compared to more complicated networks. Many of the above mention networks lack the ability to model global features as compared to local features. But in recent years by the introduction of transformers which are adopted in computer vision applications, the ability to learn global and local features of these Encoder-Decoder networks greatly increase due to the use of attention mechanisms used in these transformer-based networks.

In recent years, feature learning techniques like feature pyramids, skip connections, atrous convolutions, attention mechanism and transformers are used to enhance the feature learning process and global feature representation. Furthermore, specific or customized loss functions are used to penalize segmentation task error for histological image segmentation. However, by combining different techniques like incorporating improved attention units and specific loss function, there is room for improvement in DCNN for segmentation of histological images. To this end, we propose an Encoder-Decoder network which includes Quick Attention Unit in encoder as well as in decoder branch with a residual connection between them, and we proposed a custom loss function which is a combination of fixed focal loss, binary cross-entropy, and dice loss.

In summary, our main contributions are as follows

- Proposes a Quick Attention Unit for better global and local feature representation by avoiding irrelevant features.
- Introduces a Multi Loss Function which is a combination of fixed focal loss, binary cross-entropy loss, and dice loss which focus on learning hard negative examples and crisp boundary detection.
- Proposes an Encoder-Decoder network which capture dense multi scale features.
- Successfully improves the accuracy and performance over the different state of the art models on two different datasets.

## II. RELATED WORK

Segmentation of histological images is a task performed on complicated and small datasets for this purpose U-Net was proposed which is specially designed for segmenting tumors in the brain and lungs. authors of U-Net proposed an architecture which is an Encoder-Decoder based model in which input is downsampled using convolution layers which are performed in encoder branch and then upsampled using learnable convolution layers which are performed in the decoder branch, they also incorporated skip connections which flow the information from the encoder part to the decoder part of the network, This ability lacks in Fully Convolutional Network(FCN) [12], which improves the overall results. DCNN which are specifically designed for segmentation of histological images [16, 1, 8, 23, 20, 8, 22, 2, 6, 23, 15, 14] perform very well on medical images datasets as compared to natural images datasets. Zongwei et al proposed a U-Net++ which is variant of U-Net with modification in skip connections which merge the convolution layer in encoder with its corresponding upsampling layer in a decoder, they also incorporate dense skip connections. In MSRF-Net, authors proposed a fusion block that fuses features of multiple scales of different receptive fields, the encoder branch of MSRF-Net consists of squeeze and excitation module and the decoder branch has a triple attention mechanism that uses both channel and spatial wise attention with gated convolutions. With the advancements in image feature representation techniques, the networks are getting deeper and losing the capability of capturing relevant information. Complex and controlled attention modules are introduced to capture rich global and local contextual information, increase the receptive field of encoders, and ability to discard irrelevant information. like Prajit et al [13] incorporates self-attention units in vision models to learn feature representations, Ashish et al [20] proposed a Multi-scale self-guided attention for segmentation task in histological images.

The most recent development in computer vision are Transformers [10], they are introduced for natural language processing with more complicated attention units but these transformers can also be used for image processing tasks like in TransUNet [2], authors proposed a U shaped Encoder-Decoder network in which transformer is used as an encoder for feature extraction, the input image convert into patches and then gets into the transformer as the input sequence and a U-Net based network upsample the feature map. Jeya et al proposed a Medical Transformer [23] which incorporates gated axial attention for medical image segmentation it's a transformer-based network which have self-attention modules with control mechanism in encoder due to this strategy the dependency of pre-trained weights is removed.

## III. METHODOLOGY

Our proposed model is Encoder-Decoder network with Quick Attention Units and Multi Loss Function. We manage to perform semantic segmentation on histological images. The input of the model is RGB H&E stain image and the model predicts the binary mask in the form of a binary image as output. The whole pipeline details are discussed below:

### A. Quick Attention Unit

With the introduction of attention units in DCNN, it focuses on more relevant features of the object, so more and more complicated attention units are incorporated in DCNN which increases the computation cost so we introduced a relatively less computational expensive attention unit which is called Quick Attention. It is represented as

$$QA(x) = \sigma(f(x)^{1x1}) + x \quad (1)$$

Here x is an input feature map, $f()^{1x1}$ is a 1x1 convolution with stride 1 and the same number of filters as the input feature map, and $\sigma$ is a sigmoid function. Quick Attention takes in the feature map as an input WxHxC (Width x Height x Channels) and creates two instances of the input feature map then it performs the 1x1xC convolution on the first instance and calculates the sigmoid activations after that it is added with the second instance to generate the final attention map as output which is of same dimensions as of input. Visual representation of our proposed attention unit is in Fig 1.

### B. Encoder-Decoder

The encoder consists of layers from input to global average pooling (GAP) as shown in Fig 1 part a. After some refinements on the input layer which has a dimension of 256x256x3, using 3x3 convolution, batch normalization, and relu activations the activation map goes through a series of expanded convolution blocks to capture dense features rather than sparse features in the case of simple convolution and max-pooling operation. Each expanded convolution block consists of 1x1 convolution, batch normalization, relu activation, depthwise convolution with a residual connection between every expanded convolution as shown in Fig 1 part c. The depthwise convolution capture features at different receptive fields by specifying dilation rate parameter. After expanded convolution 6, we introduced quick attention which allows a network to capture more relevant features and from now onwards the spatial dimension of feature maps remains the same 32x32xC before global average pooling, after expanded convolutions, Atrous Spatial Pyramid Pooling (ASPP) capture features at multiple scales, it performs 1x1 convolution, 3x3 dilated convolutions with the rate 6,12,18 and image pooling after that it concatenates all the features maps and performs 1x1 convolution as shown in Fig 1 part d. After ASPP block global average pooling is performed. The decoder of the network performs upsampling and creates a 32x32x256 feature map which is then concatenated with the feature map from ASPP block, 1x1 convolution is used to match the channels of the feature map. then we again add quick attention to the decoder and create a residual connection between encoder and decoder quick attention modules, then sigmoid activation is performed followed by upsampling layer which uses bilinear interpolation by a factor of 4 and generates the maks of size 256x256x1.

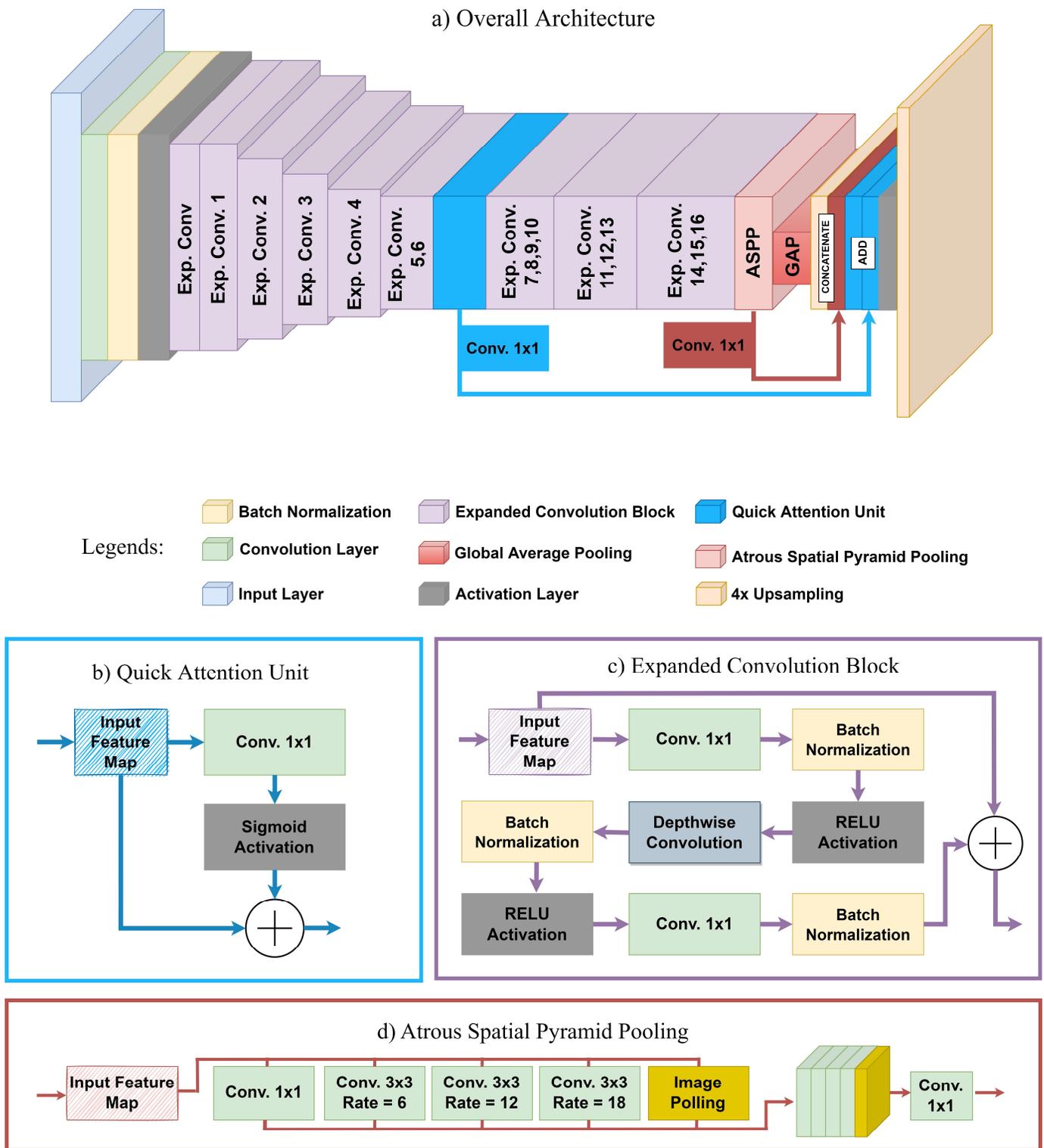

Fig. 1: Overall architecture of our proposed network, Expanded Convolutions from 1 to 6 have dilation rate = 1, from 7 to 13 have dilation rate = 2, and from 14 to 16 have dilation rate = 4, the dilation rate in expanded convolutions is used for depthwise convolution operations. Expanded Convolutions which have the same dimensions are grouped together. The overall pipeline to train the model is that we create patches of histology images using the sliding window technique and feed them into the model for training and validation. After training, we evaluate the model without creating patches full-size image is fed into the model for predictions.

## C. Multi Loss Function

Our proposed loss function is a combination of BCE Loss, Focal Loss, and Dice loss. Each one of them contributes individually to improve performance further details of loss functions are mentioned below,

(1) BCE Loss calculates probabilities and compares each actual class output with predicted probabilities which can be either 0 or 1, it is based on Bernoulli distribution loss, it is mostly used when there are only two classes are available in our case there are exactly two classes are available one is background and other is foreground. In a proposed method it is used for pixel-level classification. BCE Loss represent as

$$BCELoss(y, \bar{y}) = -(y\log(\bar{y}) + (1-y)\log(1-\bar{y})) \quad (2)$$

Here y is the actual value and $\bar{y}$ is the predicted value

(2) Focal Loss is a variant of BCE, it enables the model to focus on learning hard examples by decreasing the wights of easy examples it works well when the data is highly imbalanced, Focal Loss represent as

$$FocalLoss(p_t) = -a_t(1-p_t)^\gamma \log(p_t) \quad (3)$$

Here $\gamma$ is always greater than zero but when $\gamma$ is equal to 1 then it works like a Cross Entropy function, the range of $a$ is between 0 to 1, it is treated as a hyperparameter.

(3) Dice Loss is inspired by the Dice Coefficient Score which is an evaluation metric used to evaluate the results of image segmentation tasks. Dice Coefficient is convex in nature so it has been changed, so it can be more traceable. It is used to calculate the similarity between two images, Dice Loss represent as

$$DiceLoss(y, \bar{p}) = 1 - (2y\bar{p} + 1) \div (y + \bar{p} + 1) \quad (4)$$

Here 1 is added to ensure that function is not become undefined in edge case such as $y = p \equiv 0$

We proposed a Loss function which is a combination of all three above mention loss functions to benefit from all, BCE is used for pixel-wise classification, Focal Loss is used for learning hard examples, we use 0.25 as the value for alpha and 2.0 as the value of gamma. Dice Loss is used for learning better boundary representation, our proposed loss function represent as

$$Loss = (BCELoss + FocalLoss) + DiceLoss \quad (5)$$

## IV. DATASETS

We have performed experiments on two datasets Multi-organ Nucleus Segmentation (MoNuSeg) [11] and Gland Segmentation in Colon Histology Images (GlaS) [21], both are publicly available datasets and used for histological image segmentation task, MoNuSeg Dataset consists of 44 Hematoxylin and Eosin (H&E) stained tissue images scanned at 40x magnification. This dataset has seven organ types with benign and malignant tissue samples from multiple patients. GlaS Dataset consists of 165 H&E stained images with benign and malignant tissue samples from multiple patients.

TABLE I: Distribution of MoNuSeg dataset

| Distribution | No of images | Image Size | Mask Size |
|---|---|---|---|
| Training Set | 1470 | 256x256x3 | 256x256x1 |
| Validation Set | 686 | 256x256x3 | 256x256x1 |
| Test Set | 14 | 1000x1000x3 | 1000x1000x1 |

TABLE II: Distribution of GlaS dataset

| Distribution | No of images | Image Size | Mask Size |
|---|---|---|---|
| Training Set | 1644 | 256x256x3 | 256x256x1 |
| Validation Set | 703 | 256x256x3 | 256x256x1 |
| Test Set | 80 | 400x400x3 | 400x400x1 |

### A. Data Preprocessing

We have first created patches of each image using the sliding window technique with a 256 step size to avoid overlap. For the MoNuSeg dataset, we have generated patches from training data of 30 RGB images having a size of 1000x1000x3 (Height x Width x Channels) and their corresponding masks having a size of 1000x1000x1. For test data, we didn't apply any augmentation or create patches test data consist of 14 RGB images and their corresponding masks. We split the training data into training and validation sets, the overall dataset stats are mentioned in Table I For the GlaS dataset, we have generated patches from training data of 85 RGB images having multiple image sizes and their corresponding masks. For test data, we resize 80 images and their corresponding masks. We split the training data into training and validation sets, the overall dataset stats are mentioned in Table II.

## V. EVALUATION METRICS

To evaluate our proposed model, we employed three evaluation metrics, details are given below

(1) F1 Score is used to measure the detection accuracy of objects. When the object is segmented and if it intersects a minimum of fifty percent with its ground truth then it is considered as true positive otherwise it is a false positive. The difference between the number of objects from ground truth and the number of true positives is considered as false negatives. Given that the F1 Score represent as

$$F1(A, B) = \frac{2}{|A|/|A \cdot B| + |B|/|A \cdot B|} \quad (6)$$

Here A and B are binary vectors and the backslash is the set minus.

(2) Dice Score measures the overlap between predicted output and ground truth, so basically, it's a measure of similarity of objects. Dice Score is defined as

$$\text{Dice}(A, B) = \frac{2|A \cdot B|}{|A| + |B|} \quad (7)$$

TABLE III: Quantitative results of our proposed network

| MoNuSeg | | | GlaS | | |
|---|---|---|---|---|---|
| F1 | IoU | Dice | F1 | IoU | Dice |
| 75.08 | 71.06 | 95.20 | 98.07 | 76.73 | 99.09 |

(3) Intersection over Union (IoU) is also known as the Jaccard index, it measures the percent overlap between predicted output and ground truth. The average IoU over all classes is known as Mean IoU (mIoU). The IoU is defined as

$$\text{IoU}(A, B) = \frac{|A \cdot B|}{|\max(A, B)|} = \frac{|A|}{|A| + |B| - |A \cdot B|} \quad (8)$$

## VI. EXPERIMENTATION AND RESULTS

We performed our experiments on Google Colab. The datasets are converted into NumPy arrays after preprocessing. Train, Val, and Test splits are loaded into the main memory and fed into the network. Our proposed model is trained in end-to-end fashion for 200 epochs on MoNuSeg and for 100 epochs on GlaS, batch size of 8 was set. We use Adam optimizer with 0.01 learning rate, pre-trained mobilenetv2[17] as a feature extractor with an output stride of 8, sigmoid activations instead of softmax at the last layer, and 0.1 dropout. Quantitative results on GlaS and MoNuSeg Datasets are reported in Table III and results of comparison with other networks are reported in Table IV. Training graphs as shown in Fig 3 reveals that validation loss is decreasing after every epoch and the model does not overfit. Our proposed model outperforms the base models which shows a 1.99 % improvement on the MoNuSeg dataset and 7.15 % improvement on the GlaS dataset.The Qualitative results are shown in Fig 2.

## VII. ABLATION STUDY

For the ablation study, we incorporate different attention methods in our proposed network. The self Attention method is implemented in the encoder branch then it produces effective results but when it is implemented in the decoder branch it gives a negative F1 score and validation loss is not decreasing. Residual Channel Attention, Residual Spatial Attention, and Residual Mixed Attention (combination of Channel Attention and Spatial Attention) methods are implemented in the decoder branch and replaced by some convolution layers then they perform better otherwise when they are implemented in an encoder branch the validation loss is not decreasing and results are not improving after a few epochs. The quantitative results of different attention methods on both MoNuSeg and GlaS datasets are shown in Table V.

## VIII. CONCLUSION AND FUTURE WORK

Accurate segmentation of multi-structural objects at multiple scales in histological images is very crucial because these results are later used for further biological analysis and prognosis. To this end, we proposed a neural network which utilized Encoder-Decoder structure, Quick Attention Units, and Multi Loss Function. The proposed attention units are much simpler and effective as compared to other attention units. The attention units in our proposed method are used in both encoder and decoder with residual connection to ensure that model focuses on more relevant global and local features at multiple scales. Quick Attention Units combined with Multi Loss Function produces more accurate and improved results over multiple datasets which shows the improved generalization capability of our proposed model. In future we extend the model capability to capture multi class and multi structure features in a single pipeline, include more datasets to evaluate the generalization capability of model, build a pipeline to use our proposed approach in instance segmentation and panoptic segmentation area and introduce model inference optimization by quantization of weights and skip some layers so we can decrease the model inference time and use it on edge devices.

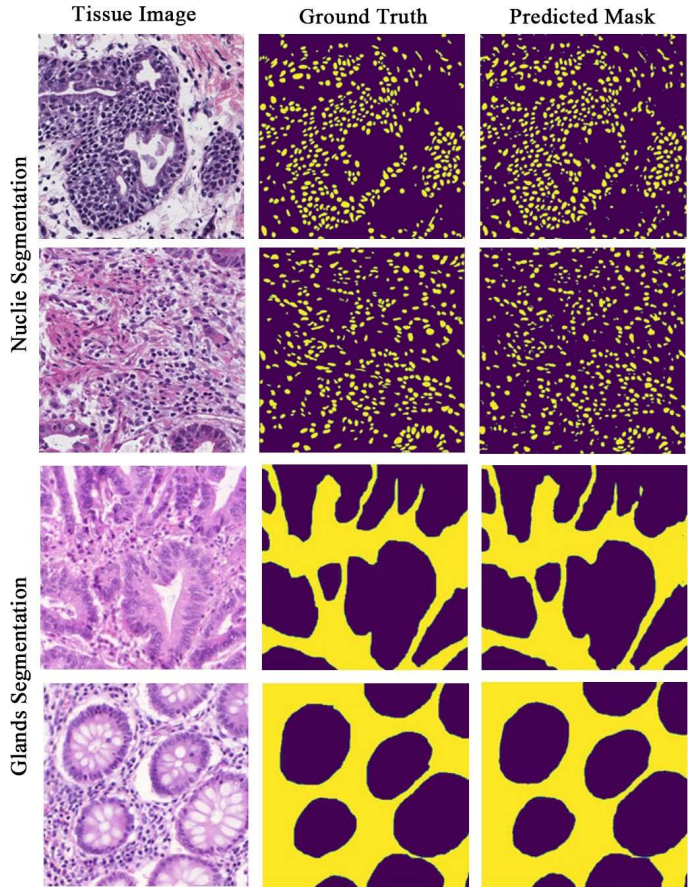

Fig. 2: Qualitative results of our proposed method.First two rows shows the results from MoNuSeg Dataset and last two rows shows the results from GlaS Dataset.

TABLE IV: Quantitative comparison of the proposed method with other networks

| Network | MoNuSeg | | GlaS | |
|---|---|---|---|---|
| | F1 | IoU | F1 | IoU |
| FCN [12] | 28.84 | 28.71 | 66.61 | 50.84 |
| U-Net [16] | 79.43 | 65.99 | 77.78 | 65.34 |
| U-Net++ [27] | 79.49 | 66.04 | 78.03 | 65.55 |
| Res-UNet [26] | 79.49 | 66.07 | 78.83 | 65.95 |
| Deeplabv3+ [3] | 72.16 | 66.56 | 76.01 | 67.04 |
| Axial Attention U-Net [25] | 76.83 | 62.49 | 76.26 | 63.03 |
| MedT [9] | **79.55** | 66.17 | 81.02 | 69.61 |
| Proposed Network | 75.08 | **71.06** | **98.07** | **76.73** |

TABLE V: Ablation Study

| Attention Method | MoNuSeg | GlaS |
|---|---|---|
| Self Attention [13] | 0.56 IoU | 0.61 IoU |
| Residual Channel Attention [24] | 0.37 IoU | 0.31 IoU |
| Residual Spatial Attention [24] | 0.37 IoU | 0.35 IoU |
| Residual Mixed Attention [24] | 0.40 IoU | 0.45 IoU |

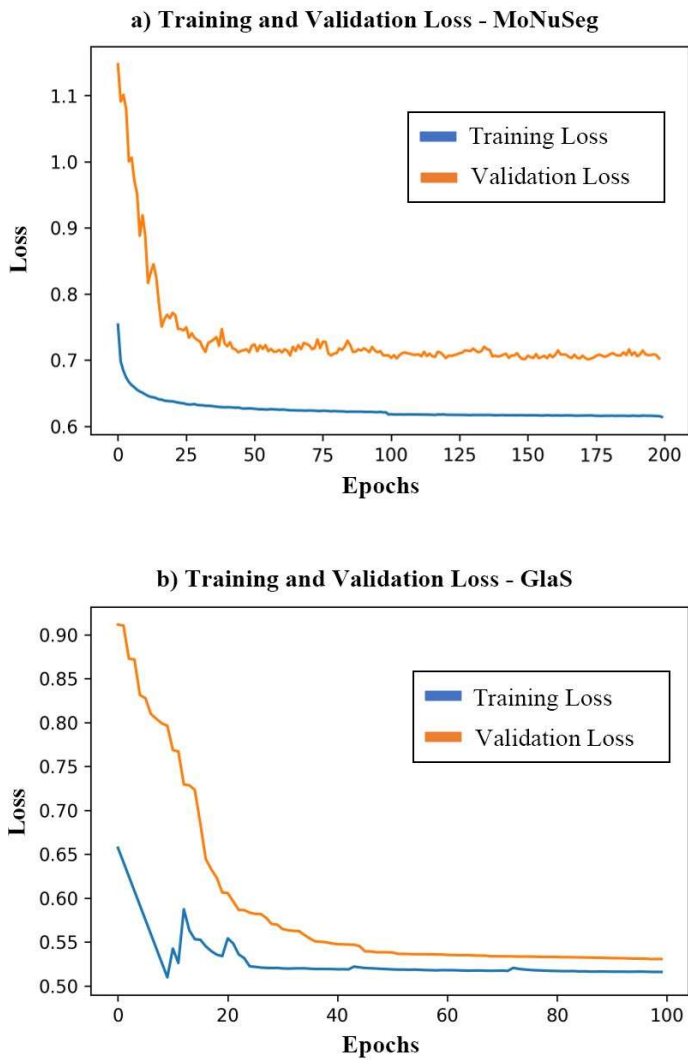

Fig. 3: Training and Validation Loss graphs of our proposed network